\documentclass[10pt,preprintnumbers,amsmath,amssymb,floatfix,twocolumn,prl]{revtex4-2}
\usepackage[utf8]{inputenc}
\usepackage{lipsum, babel}
\usepackage{mathrsfs}
\usepackage{graphics}
\usepackage{xspace}
\usepackage[T1]{fontenc}
\usepackage{graphicx}
\usepackage{dcolumn}
\usepackage{bm}
\usepackage{float}
\usepackage{verbatim}
\usepackage[utf8]{inputenc}
\usepackage{graphicx}
\usepackage{bm, physics, slashed}
\usepackage[dvipsnames]{xcolor}
\usepackage[colorlinks=true]{hyperref}
\usepackage[capitalise]{cleveref}

\begin{document}

\title{Emergent Error Correcting States in Networks of Nonlinear Oscillators}

\author{X. Jin}
\author{C. G. Baker}
\author{E. Romero}
\author{N. P. Mauranyapin}
\author{T. M. F. Hirsch}
\author{W. P. Bowen}
\email[Corresponding author: ]{w.bowen@uq.edu.au}
\author{G. I. Harris}
\affiliation{School of Mathematics and Physics, The University of Queensland, QLD 4072, Australia}

\begin{abstract} 

Networks of nonlinear oscillators can exhibit complex collective behaviour ranging from synchronised states to chaos. Here, we simulate the dynamics of three coupled Duffing oscillators whose multiple equilibrium states can be used for information processing and storage. Our analysis reveals that even for this small network, there is the emergence of an \emph{error correcting phase} where the system autonomously corrects errors from random impulses. The system has several surprising and attractive features, including dynamic isolation of resonators exposed to extreme impulses and the ability to correct simultaneous errors. The existence of an error correcting phase opens the prospect of fault-tolerant information storage, with particular applications in nanomechanical computing.

\end{abstract}

\maketitle

Networked nonlinear oscillators often display intricate emergent phenomena that defy prediction from extrapolation of the behaviours of individual oscillators. A prime example is when a network of unsynchronized oscillators spontaneously achieves synchronization once their coupling strength surpasses a critical threshold~\cite{Restrepo_PRE_05}. Adjusting system parameters can further lead to the formation of chimera states~\cite{Abrams_PRL_04}, where various subsets of oscillators exhibit distinct dynamic patterns, some of which may even evolve into chaotic behavior. Understanding of these intricate behaviors has yielded valuable insights into a diverse range of physical occurrences, such as the synchronized flashing of fireflies~\cite{Sarfati_SciAdv_22}, the neurological underpinnings of Parkinson's disease~\cite{Uhlhass_Neuron_06}, and the emergence of novel phases of matter in quantum systems~\cite{Sakurai_PRL_21}. They can also be utilised in technological applications, such as clocks with improved timing uncertainty~\cite{Komar_NatPhys_2014}, neuromorphic computing~\cite{Torrejon_Nat_2017}, and sensing networks~\cite{Xu_MicroNano_2020}. 
    
Many physical systems exhibit nonlinear oscillations, such as superconducting circuits~\cite{Boaknin}, optical parametric oscillators/amplifiers~\cite{Moss}, and spin-waves in magnetic materials~\cite{Woltersdorf_NatCom_2022}. One particularly important example is the nanomechanical oscillator, which has played a critical role in the development of modern technologies, with applications ranging from ultra-precise chronometry~\cite{gavartin2013stabilization} and sensing~\cite{zhang2005application} to filtering~\cite{delsing_JPhysD_2019} and imaging~\cite{rutzel_RoyalSoc_2003}. 
Nanomechanical oscillators have also been identified as a promising platform for alternative forms of computation, enabling efficient and robust operation in harsh environments, such as those in medical facilities, nuclear power plants, and outer space~\cite{Coulombe_PLOSONE_2017, Song_19, Wenzler, Yamaguchi_11, Yao_14, Roukes}.

Here, we explore the emergent behaviour of simple networks of three all-to-all coupled oscillators. We include a third-order nonlinearity; the dominant nonlinearity for many oscillators, and of particular relevance in nanomechanical systems where it is referred to as a Duffing nonlinearity. The characteristic bistability of each Duffing oscillator is used to define logical states for information processing~\cite{Romero_arxiv22}. We hypothesise that, in appropriate parameter regimes, the coupling may act as a stabiliser correcting for logical errors in individual oscillators. A computational search over the parameter space of the system confirms this hypothesis, showing that over a wide region, a collective error correcting phase emerges. That is, the proposed system returns to its initial state after being disturbed by an instantaneous perturbation. The system has several striking features that distinguish it from a traditional 3-bit majority-vote error correction scheme, including autonomous operation, enhanced performance for extreme perturbations, and the ability to probabilistically correct simultaneous errors.  

\begin{figure*}[ht!]
\begin{center}
\scalebox{0.5}{\includegraphics{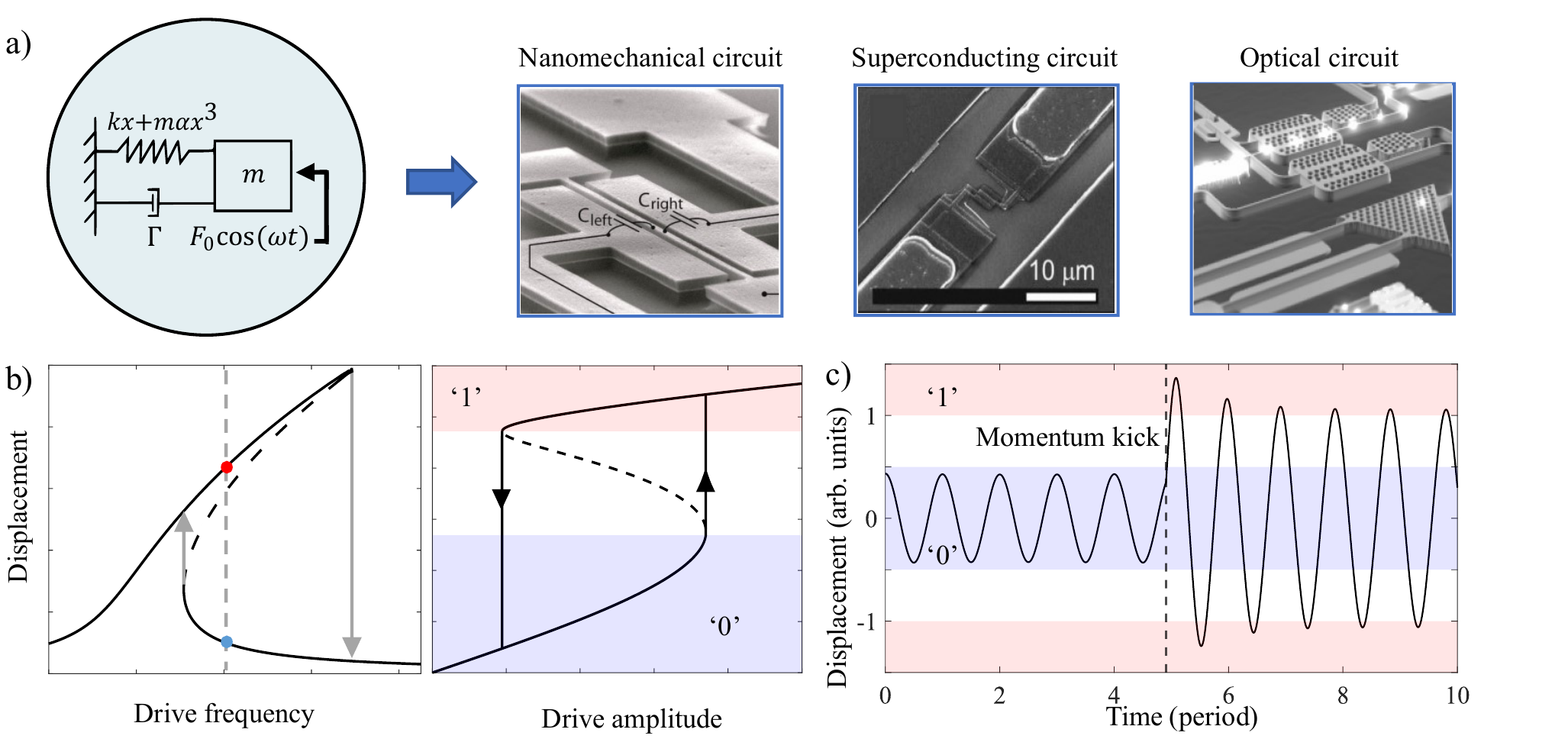}}
\end{center}
\caption{a) Schematic of a single Duffing oscillator and its applications in various computing platforms. Images of the nanomechanical circuit, superconducting circuit, and optical circuit reproduced from Ref~\cite{Wenzler,Boaknin,Moss}, respectively. b) Frequency/drive response of Duffing oscillator. Solid (dashed) lines represent stable (unstable) solutions of the Duffing equation. Jumps between the stable solutions are labelled with up and down arrows. c) Time dynamics of a single Duffing oscillator subject to an SEU. The parameters of the oscillator are given by: $m = 10^{-12}$ kg, $\Gamma = 10^{5}$ s$^{-1}$, $\omega_0 = 10^{6}$ s$^{-1}$, $\alpha = 2 \times 10^{22}$ m$^{-2} s^{-2}$, $\omega = 1.152 \times 10^{6}$ s$^{-1}$, $F  = 5 \times 10^{-7}$ N, $\Delta p = 6 \times 10^{-12}$ kg.m.s$^{-1}$. The oscillator is first initialised in the `0' state, but then subject to an SEU which leads to a bit flip. The displacement is expressed in arbitrary units, such that the magnitude of a `1' signal is $1$.}
\label{Figure 1}
\end{figure*}

Despite their robustness to environmental influence, nanomechanical logic elements are susceptible to transient errors such as those arising from thermal effects, electromagnetic pulses, ionising radiation, or mechanical shock~\cite{Lee_NatComm_23}. These types of transient bit-errors, which also occur frequently in conventional computers~\cite{schroeder_Google_2009}, are often referred to as single-event upsets (SEUs)~\cite{Karnik_IEEE_04, Ziegler_Science_79}. For nanomechanical logic, SEUs originate from an instantaneous impulse on the mechanical oscillator that is sufficiently large to shift the oscillator from one logical state to the other. Our discovery of an error-correcting phase, which applies broadly to third-order nonlinear oscillators, provides a means to address these errors, paving the way towards fault-tolerant nanomechanical computing.

The Duffing oscillator is one of the most widely studied nonlinear systems~\cite{Yao, Tadokoro21}, with some common applications illustrated in Fig~\ref{Figure 1}a). It is described by the following equation~\cite{Schmid}:
\begin{align} 
\ddot{x} + \Gamma \dot{x} + \omega_{0}^{2} x + \alpha x^3 &= \frac{F }{m} \cos(\omega t),\label{Duffing}
\end{align}
where, for a mechanical oscillator, $x$ is the displacement, $m$ the mass, $\Gamma$ the dissipation, $\omega_0$  the resonant frequency, $\alpha$ the strength of nonlinearity and $F$ the amplitude of the sinusoidal drive provided to the system at frequency $\omega$. 
At certain drive amplitudes and frequencies, the steady-state dynamics exhibit bistability~\cite{Schmid}, with one stable solution corresponding to high displacement and the other corresponding to low displacement~\cite{Romero_arxiv22}. This bistability is shown in Fig.~\ref{Figure 1}b) for the case of spring hardening ($\alpha$ >0), wherein the oscillation amplitude depends on the history of the drive frequency (left) and amplitude (right).

We use numerical methods to simulate the responses of Duffing oscillators to incident impulses, which are modelled as instantaneous changes to the momentum of the oscillators. By convention, we use high (low) displacements to represent binary `1' (`0') signals (red/blue shaded regions in Fig.~\ref{Figure 1}b))~\cite{Tadokoro21, Romero_arxiv22}. Figure~\ref{Figure 1}c) illustrates the time dynamics of a single Duffing oscillator based on Eq.~\ref{Duffing}, with an impulse introduced at the time indicated by the dashed line. The oscillator is initially prepared in a `0' state but the impulse causes it to latch onto the `1' state, which corresponds to an erroneous bit-flip.

\begin{figure*}[hbt!]
\begin{center}
\scalebox{0.55}{\includegraphics{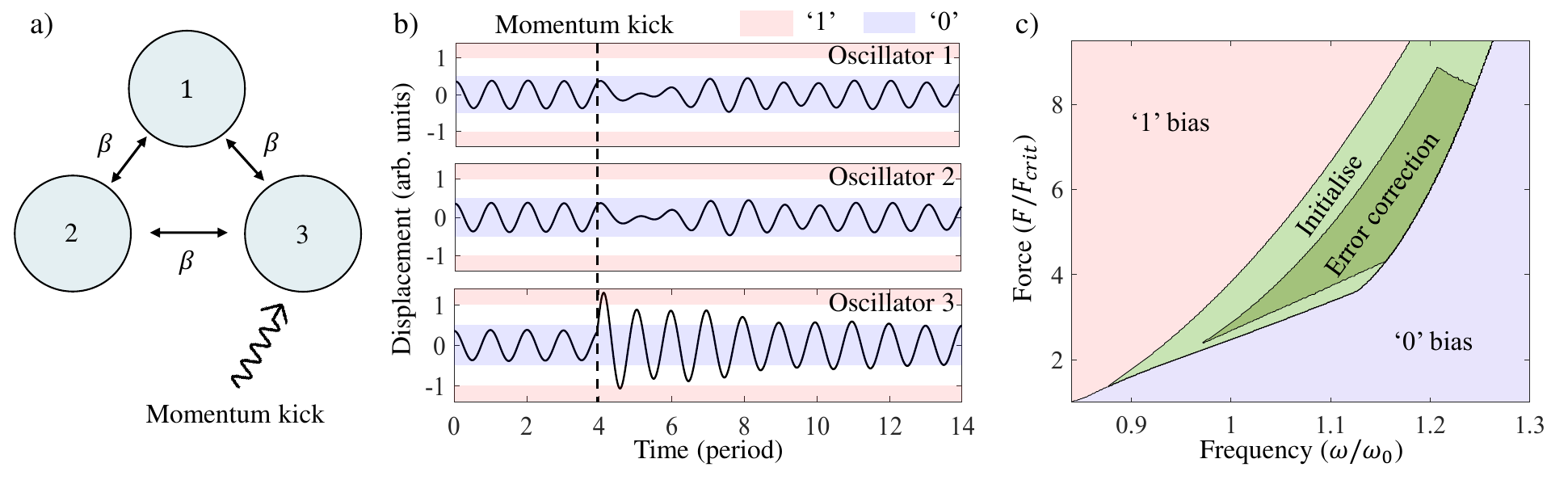}}
\end{center}
\caption{a) Schematic of mechanical error correction system, composed of three coupled Duffing oscillators. This functions as a majority voting system and can correct for single SEUs. b) Time dynamics of error correction device. The coupled oscillators are initialised in their `0' states and after 4 periods of oscillation, an impulse is applied to the third oscillator. The third oscillator temporarily transitions into its `1' state, but quickly equilibrates back.  The parameters of the oscillator are given by: $m = 10^{-12}$ kg, $\Gamma = 10^{5}$ s$^{-1}$, $\omega_0 = 10^{6}$ s$^{-1}$ , $\alpha = 2 \times 10^{22}$ m$^{-2}.s^{-2}$, $\omega = 1.152 \times 10^{6}$ s$^{-1}$, $F=1.23 \times 10^{-6}$ N, $\Delta p = 6 \times 10^{-12}$ kg.m.s$^{-1}$, $\beta = 2 \times 10^{11}$ s$^{-2}$. c) Phase map of error correction device. The map is divided into four main regions, `1' bias, `0' bias, initialise and error correction. This map is produced using the same parameters as b).} 
\label{Figure 2}
\end{figure*}

In conventional computing systems, a common technique to mitigate the effects of SEUs is to use majority voting logic ~\cite{TMR}. Within a 3-bit majority voting algorithm, each bit is replicated three times; if one bit is affected by a SEU, then the algorithm corrects this error by assuming that the majority state of the 3 bits is correct. This is traditionally achieved through bit post-processing with a logic circuit. Here, we design a similar scheme in nanomechanical logic, in which the error correction occurs autonomously, without the need for additional comparator circuitry. 
We consider three identical, equally forced, Duffing oscillators that are all linearly coupled (see Fig.~\ref{Figure 2}a)). Each individual oscillator is driven into the bistable region and represents a single logical bit. The coupling provides an additional avenue for driving that can be either large or small depending on the amplitude of the neighbouring oscillators. The equations of motion of these coupled oscillators are:
\begin{align}
\ddot{x}_1 + \Gamma \dot{x}_1 + \omega_{0}^2 x_1 + \alpha x_1^{3}& = \frac{F}{m} \cos(\omega t) + \beta x_2 + \beta x_3 \label{coupled_one}  \\
\ddot{x}_2 + \Gamma \dot{x}_2 + \omega_0^2 x_2 + \alpha x_2^{3}  & = \frac{F}{m} \cos(\omega t)  + \beta x_1 + \beta x_3 \label{coupled_two} \\
\ddot{x}_3 + \Gamma \dot{x}_3 + \omega_0^2 x_3+ \alpha x_3^{3}  &= \frac{F }{m}   \cos(\omega t)  + \beta x_1 + \beta x_2  \label{coupled_three} 
\end{align} 
where $\beta$ is the coupling rate between oscillators, and $x_1$, $x_2$, and $x_3$ represent the displacement of oscillators $1$, $2$ and $3$, respectively.

To determine the collective behaviour of the proposed error correcting network, we simulate its time dynamics for a range of different parameters and initial conditions. We search for emergent phases where the collective dynamics exhibit only two steady states ('111' and '000'), irrespective of when impulses occur and insensitive to small deviations in system parameters. We find that for specific sets of parameters, error correcting behaviour is possible. One example is shown in Fig.~\ref{Figure 2}b), where each oscillator is initially prepared in the `0' state, before oscillator 3 is subjected to an impulse of the same magnitude as that in Fig.~\ref{Figure 1}c) (dashed line). We see that the impulse causes the amplitude of oscillator 3 to temporarily reach the `1' state, however, it does not latch to this state. After an equilibration period, all oscillators settle back into the `0' state. Similar error correcting dynamics are seen when the oscillators are initialised in the '1' states (see Supplementary Information).

Next, we repeat these simulations varying the drive frequency, drive amplitude and time of impulse. We find the behaviour of the coupled system can be categorized into four distinct phases (see Fig.~\ref{Figure 2}c)). When the drive is strong and near resonance, all oscillators evolve into their `1' states. Conversely, when the oscillators are provided with weak drives, or are driven far away from resonance, they always evolve into their `0' states. We term these two regions \emph{`1' bias} and \emph{`0' bias}, respectively. In between the two bias regions, the oscillators can be initialized in either their collective `1' or `0' states, depending on the history of the applied forcing. We call this the \emph{initialise} region. It should be noted that operating in this region does not guarantee error correction. We find that, for some parameters and impulse timings, it is possible for a single impulse to cause the system to transition between collective states. However, there remains a subspace that is guaranteed to correct all single impulses. This region is labelled \emph{error correction} in Fig.~\ref{Figure 2}c).

For sufficiently high coupling rates, the error correction region can be made to be large, enabling stable operation. For instance, Fig.~\ref{Figure 2}c) was generated with a coupling rate equal to twice the intrinsic dissipation (i.e. $\beta/\omega_0=2\Gamma$). Here, near the centre of the error correction region, error correction is maintained even with large deviations in drive force. For instance, at $\omega/\omega_0 = 1.152$, error correction is predicted to occur for drive forces in the range $4.3 F_{crit} \leq F \leq 6.8 F_{crit}$. The size of the error correction region decreases with decreasing coupling rate, eventually disappearing altogether in the weak-coupling regime where $\beta/\omega_0 <\Gamma$.

. 

\begin{figure}[hbt!]
\begin{center}
\includegraphics[width=1.0 \columnwidth]{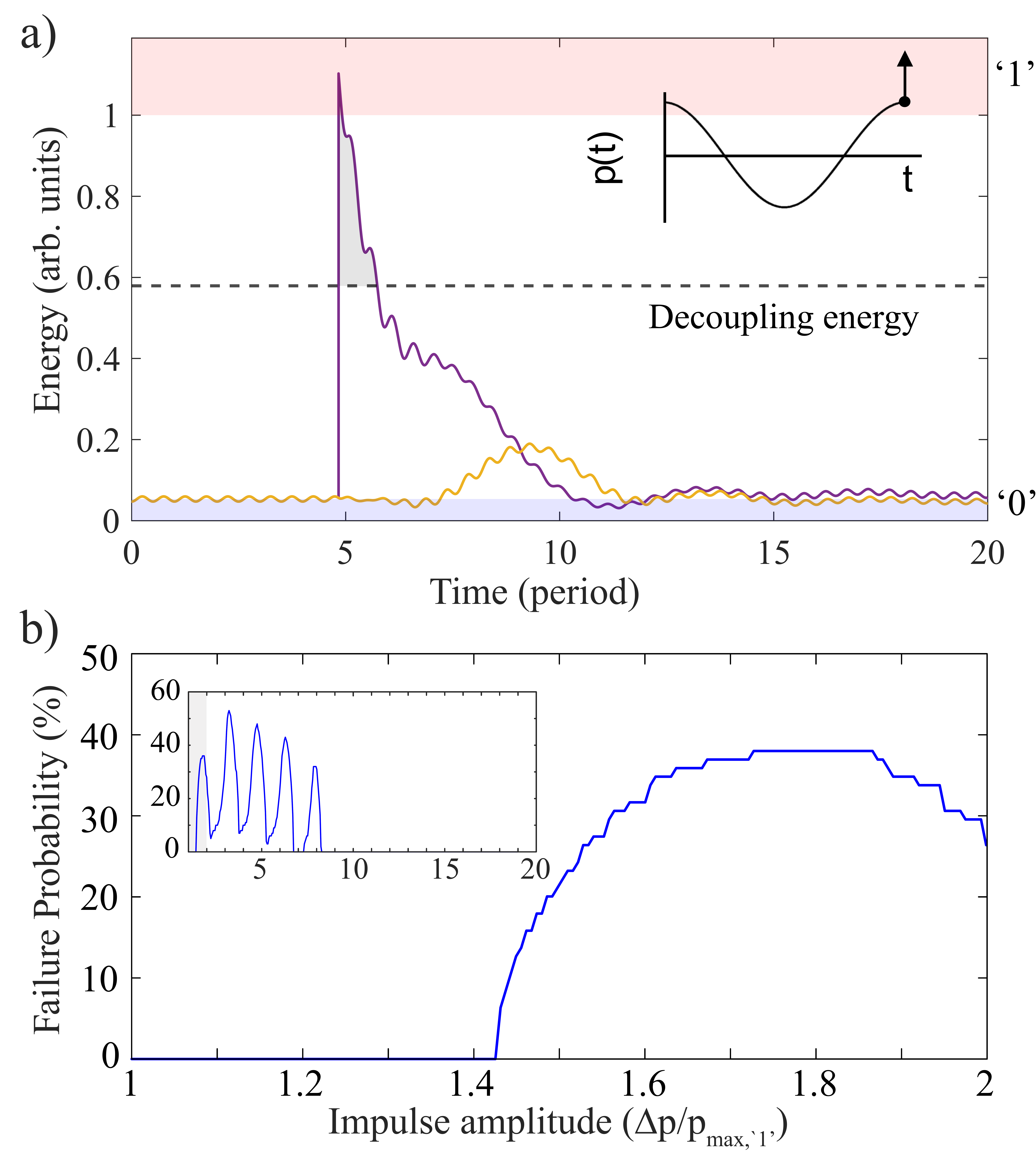}
\end{center}
\caption{Single event upset on a 3-bit majority system. a) Instantaneous energy of coupled system. Oscillators 1 and 2 are represented by the yellow trajectory and oscillator 3 is represented by the purple trajectory. The phase of the impulse relative to motion of the oscillator is indicated by the diagram on the top right corner of the figure. Since the amplitude of oscillator 3 is momentarily increased, its resonance frequency is up-shifted and it decouples from the other oscillators. The energy required for this to occur is indicated by a horizontal dashed line, and the region of decoupling is represented by grey shading. b) Probability of error-correction failure (i.e. error occurs from impulse) with increasing amplitude of impulse. The horizontal axis is normalised to the maximum momentum of the oscillator when in the `1' state. Inset: probability of error-correction failure for a larger range of impulse amplitudes. Grey region is the range of the main figure. Note that at high impulse amplitudes (i.e. $\Delta p > 8.5  p_{\text{max,`1'}}$) the error-correction is always successful.
}
\label{Figure 3}
\end{figure}

The error correction of all single impulses can be further understood by considering the instantaneous energy of the system as a function of time. Figure~\ref{Figure 3}a) shows that the energy of the perturbed oscillator (purple line) increases rapidly due to the impulse, as expected. 
 
The two unperturbed oscillators then extract the excess energy and dissipate it to the environment, bringing the perturbed oscillator back into its original state. As the impulse amplitude is increased, the ability of the unperturbed oscillators to dissipate the excess energy is reduced, causing failures in error correction. This is shown in Fig.~\ref{Figure 3}b) where impulses that cause the system momentum to exceed $p> 1.42 p_{\text{max,`1'}}$ result in failures of error correction (where $p_{\text{max,`1'}}$ is the maximum momentum of the equilibrated `1' state).

Unexpectedly, we find that the system also exhibits error correction of extremely large impulses (i.e. impulse energy large enough to transition all oscillators into the error state). This occurs due to an effect we call \textit{dynamic decoupling}. 
An extremely large impulse will temporarily increase the amplitude of the perturbed oscillator, nonlinearly shifting its oscillation frequency away from the remaining oscillators. This decouples the perturbed oscillator from the others and precludes efficient energy exchange between them. The perturbed oscillator then dissipates the excess energy to the environment before reducing its amplitude and re-establishing coupling.
The dashed line in Fig~\ref{Figure 3}a) indicates the approximate energy threshold above which dynamic decoupling starts to occur, defined as the point when the Duffing induced frequency shift exceeds the coupling rate between oscillators $~\beta/\omega_0$ (See Supplemental Information). Indeed, dynamic decoupling becomes more pronounced for larger impulse amplitudes. As a result, for our set of parameters, there are two regimes that allow perfect error correction of SEUs; impulse amplitudes in the range $0<\Delta p/p_{\text{max,`1'}}<1.42$ and those with $\Delta p/p_{\text{max,`1'}}>8.5$. This is shown in Fig.~\ref{Figure 3}b) where the failure probability is zero for both small (main figure) and large (inset) impulse amplitudes.

The error correction protocol can perfectly correct single impulses, but we expect it to be susceptible to multiple simultaneous impulses. To explore this, we run simulations in which each oscillator has a fixed probability, $P_{kick}$, of experiencing an impulse at a given time. Consequently, there is a finite probability of two or all three oscillators experiencing impulses simultaneously. We define an event as the occurrence of one or more impulses. If an event occurs and causes an error, then it is recorded and the simulation is reinitialised (see Supplementary Information for details). This allows one to determine the probability of failure for a given event probability. When this process is repeated for a range of event probabilities, the probability of failure (normalised by the event probability itself) is obtained.

\begin{figure}[hbt]
\begin{center}
\includegraphics[width=.9 \columnwidth]{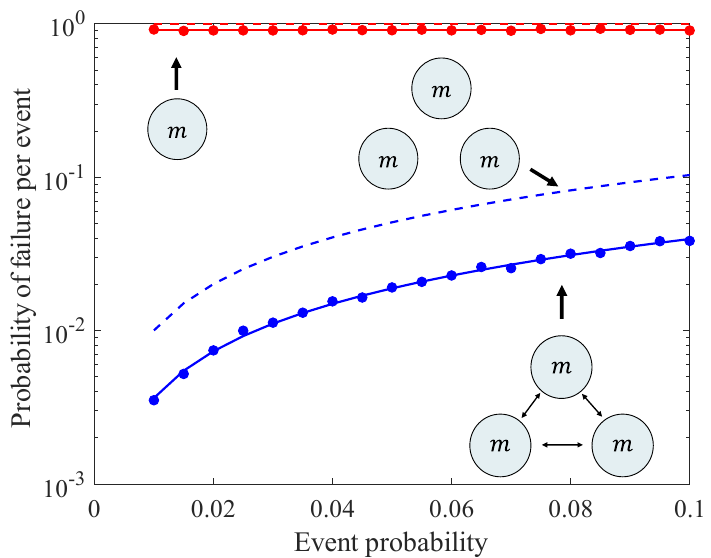}
\end{center}
\caption{Simulated probability of system failure per event as a function of event probability per time interval. Red and blue dots represent the simulation results of single and coupled systems, respectively. Red and blue dashed lines represent the predicted model for single oscillator and majority vote of three independent oscillators, respectively. Red and blue solid lines represent the corrected model for single oscillator and coupled oscillators, respectively. Note: simulation parameters are the same as in Figure 2.   }
\label{Probablity Model}
\end{figure}

We compare to equivalent simulations on a single oscillator (red in Fig.~\ref{Probablity Model}).
The probability of failure per event for the single oscillator is constant at approximately 91\%. The $9 \%$ success rate can be attributed to a portion of impulses opposing the momentum of the oscillator and therefore having a reduced effect.

The performance of the system was investigated with respect to an ideal majority voting scheme. Three way majority voting logic should correct all single impulses, but fail if two or more oscillators experience them. One can predict the theoretical failure rate, so defined, from a Binomial distribution for a given impulse probability (see Supplementary Information). 
When comparing the three oscillator system (blue dots) to the theoretical performance of a conventional majority voting scheme (blue dashed line), Fig.~\ref{Probablity Model} shows that the coupled system surprisingly exhibits superior error-correcting performance.

Interestingly, the coupled system corrects all single impulses \emph{and} 65\% of the double impulses. We believe that the ability of the coupled oscillators to correct a double impulse is related to non-trivial transients of the system. Quite generally, as the oscillator transitions between its two stable states, it must re-phase its oscillation with respect to the external drive. In this situation, the displacement from an impulse may not have the correct phase to enable latching onto the stable state, causing the impulse energy to be extracted from the system through dissipation and the external drive. In some sense, the state of each oscillator is encoded twice in the dynamics, once in its amplitude and again in its phase. This creates additional redundancy of the logical bit, enabling double errors to be corrected.

In this letter, we have explored the emergence of an error correcting phase in systems of coupled nonlinear oscillators. Using numerical simulations, we identify four distinct regions of behaviour in parameter space. We show that one of these regions enables autonomous error-correction from randomly occurring impulses, removing the need for additional logic operations to perform a majority-vote. Interestingly, error-correction is greatly enhanced for large impulses due to a \textit{dynamic decoupling} of the perturbed oscillator, enabling error correction for impulses that are over ten times larger than the momentum of the `1' state. Furthermore, the system is capable of correcting two simultaneous errors 65\% of the time due to transient effects that de-phase the oscillators with respect to the drive.

Our work shows that the complexity of networks of nonlinear oscillators can be harnessed to perform useful computational tasks, and potentially enable fault-tolerant nanomechanical computing.

This research was primarily funded by the Australian Research Council and Lockheed Martin Corporation through the Australian Research Council Linkage Grant No. LP190101159. Support was also provided by the Australian Research Council Centre of Excellence for Engineered Quantum Systems (No. CE170100009). G.I.H. and C.G.B. acknowledge their Australian Research Council grants (No. DE210100848 and No. DE190100318), respectively.

\bibliographystyle{apsrev4-2}

\providecommand{\noopsort}[1]{}\providecommand{\singleletter}[1]{#1}%
\newpage 

\clearpage
\renewcommand{\thesection}{\Alph{section}}
\renewcommand{\thesubsection}{\thesection.\arabic{subsection}}
\setcounter{section}{0}
\onecolumngrid
\section*{Supplemental Material}

\section{Strong Coupling Threshold of Three Coupled Harmonic Oscillators}

The equations of motion for three coupled harmonic oscillators are given by:

\begin{align}
\ddot{x}_1 + \omega_0^2 x_1 & = \beta x_2 + \beta x_3   \\
\ddot{x}_2 + \omega_0^2 x_2  & = \beta x_1 + \beta x_3 \\
\ddot{x}_3  + \omega_0^2 x_3  &=  \beta x_1 + \beta x_2  
\end{align} where $\omega_0$ is the resonant frequency of an uncoupled oscillator and $\beta$ is the coupling strength. Assuming that the solution takes the form of $x_j (t) = x_{j}^0 \exp(- i \omega t)$, the equations of motion can be rewritten in matrix form:
\begin{align*}
\boldsymbol{A} \ x & = \boldsymbol{0}\\
\begin{bmatrix}
-\omega^2 + \omega_0^2 &  - \beta & -\beta \\
 - \beta & -\omega^2 + \omega_0^2 & -\beta \\
  - \beta &  -\beta & -\omega^2 + \omega_0^2 
\end{bmatrix} \begin{bmatrix}
x_1^0 \\
x_2^0 \\
x_3^0
\end{bmatrix} & = \begin{bmatrix}
0\\
0\\
0
\end{bmatrix}
\end{align*}
The characteristic equation of the matrix $\boldsymbol{A}$ is given by:
\begin{align*}
- 2 \beta^3 + 3 \beta^2  \omega^2 - \omega^6 - 3 \beta^2 \omega_0^2 + 3 \omega^4 \omega_0^2 - 3 \omega_0^2 \omega^4 + \omega_0^6 &= 0
\end{align*}
By solving for  $\omega^2$, one obtains:
\begin{align*}
\omega_1 ^2 &= \omega_0^2 - 2 \beta\\
\omega_2^2 &= \omega_0^2 + \beta \\
\omega_3^2 &= \omega_0^2 + \beta
\end{align*}

The above expressions describe the frequencies of the three fundamental modes of oscillation of the system.  
The mode utilised by the proposed error correction device was determined to be the lowest frequency mode. The frequency shift introduced by coupling for this particular mode is given by:
\begin{align*}
\omega_1  &=\sqrt{\omega_0^2 - 2 \beta} \\
 &\approx \omega_0 - \frac{\beta}{\omega_0}
\end{align*}
For the oscillators to be strongly coupled, the frequency shift must be greater than the dissipation rate ($\Gamma$, as defined in the main text):
\begin{align*}
\omega_0 - \omega_1 \approx \frac{\beta}{\omega_0} \geq \Gamma
\end{align*}


\section{Decoupling energy of Three Coupled Duffing Oscillators}

In comparison, the equations of motion of three damped and driven coupled Duffing oscillators with mass '$m$' and damping rate '$\gamma$' are given by:
\begin{align}
\ddot{x}_1 + \gamma \dot{x}_1 + \omega_0^2 x_1 + \alpha x_1^{\ 3}& = \frac{F}{m} \cos(\omega t) + \beta x_2 + \beta x_3 \label{Supp coupled_one}  \\
\ddot{x}_2 + \gamma \dot{x}_2 + \omega_0^2 x_2 + \alpha x_2^{\ 3}  & = \frac{F}{m} \cos(\omega t)  + \beta x_1 + \beta x_3 \label{Supp coupled_two} \\
\ddot{x}_3 + \gamma \dot{x}_3 + \omega_0^2 x_3+ \alpha x_3^{\ 3}  &= \frac{F }{m}   \cos(\omega t)  + \beta x_1 + \beta x_2  \label{Supp coupled_three} 
\end{align}
For a single Duffing oscillator, the shift in resonant frequency as a function of displacement is given by the backbone equation (\cite{Wawrzynski_2021, BrennanM_2018}):
\begin{align}
    \omega_{backbone} = \sqrt{\omega_0^2 (1+\frac{3 \alpha}{4 \omega_0^2} x^2)}
\end{align}
where $\omega_0$ is the resonant frequency of a single oscillator and $\alpha$ is the Duffing coefficient, which quantifies the degree of non-linearity. We consider the condition of decoupling to be where the frequency shift induced by the kick is greater than that required to sustain strong coupling (as defined in the previous section). This can be written as:
\begin{align}
\omega_d - \omega &\geq \beta/\omega_0
\end{align}
where $\omega_d$ is resonance frequency required for decoupling. That is, when comparing to the other oscillators, the shift in resonance frequency of the kicked oscillator is greater or equal to the coupling rate. Assuming all the oscillators are initialised their `0' states with steady state displacement $x_0$, the displacement required for decoupling, $x_d$, is then given by: \\
\begin{align*}
\sqrt{\omega_0^2 + \frac{3 }{4 }  \alpha  x_d ^2} - \sqrt{\omega_0^2 + \frac{3 }{4 } \alpha  x_0 ^2} &= \beta/\omega_0\\
\omega_0^2 + \frac{3 }{4 } \alpha x_d^2 &= \left( \beta/\omega_0 + \sqrt{\omega_0^2 + \frac{3 }{4 }  \alpha  x_0 ^2} \  \right)^2\\
x_d^2 &= \frac{4}{3 \alpha} \left[ \left(\beta/\omega_0 + \sqrt{\omega_0^2 + \frac{3}{4} \alpha  x_0 ^2} \right)^2 - \omega_0^2 \right]
\end{align*}

The energy of each oscillator is given by the sum of its potential and kinetic energy:\\
\begin{align*}
E& = PE + KE\\
&= \frac{1}{2} kx^2 + \frac{1}{4} m \alpha x^4 + \frac{1}{2} mv^2
\end{align*}
Here, $k$ is the natural spring constant, given by $m \omega_0 ^2$, and $x_{d}$ is the steady state displacement required for decoupling. When the oscillator reaches its maximum displacement, it attains its maximum potential energy, but has zero kinetic energy. Therefore, we can calculate the energy required for decoupling by calculating the potential energy at $x_{d}$:
\begin{align*}
E_{d} &= \frac{1}{2} kx_{d}^2 + \frac{1}{4} m \alpha x_{d}^4
\end{align*}


\section{Error correction of `111' state}

\begin{figure}[hbt!]
\begin{center}
\includegraphics[width=.5 \columnwidth]{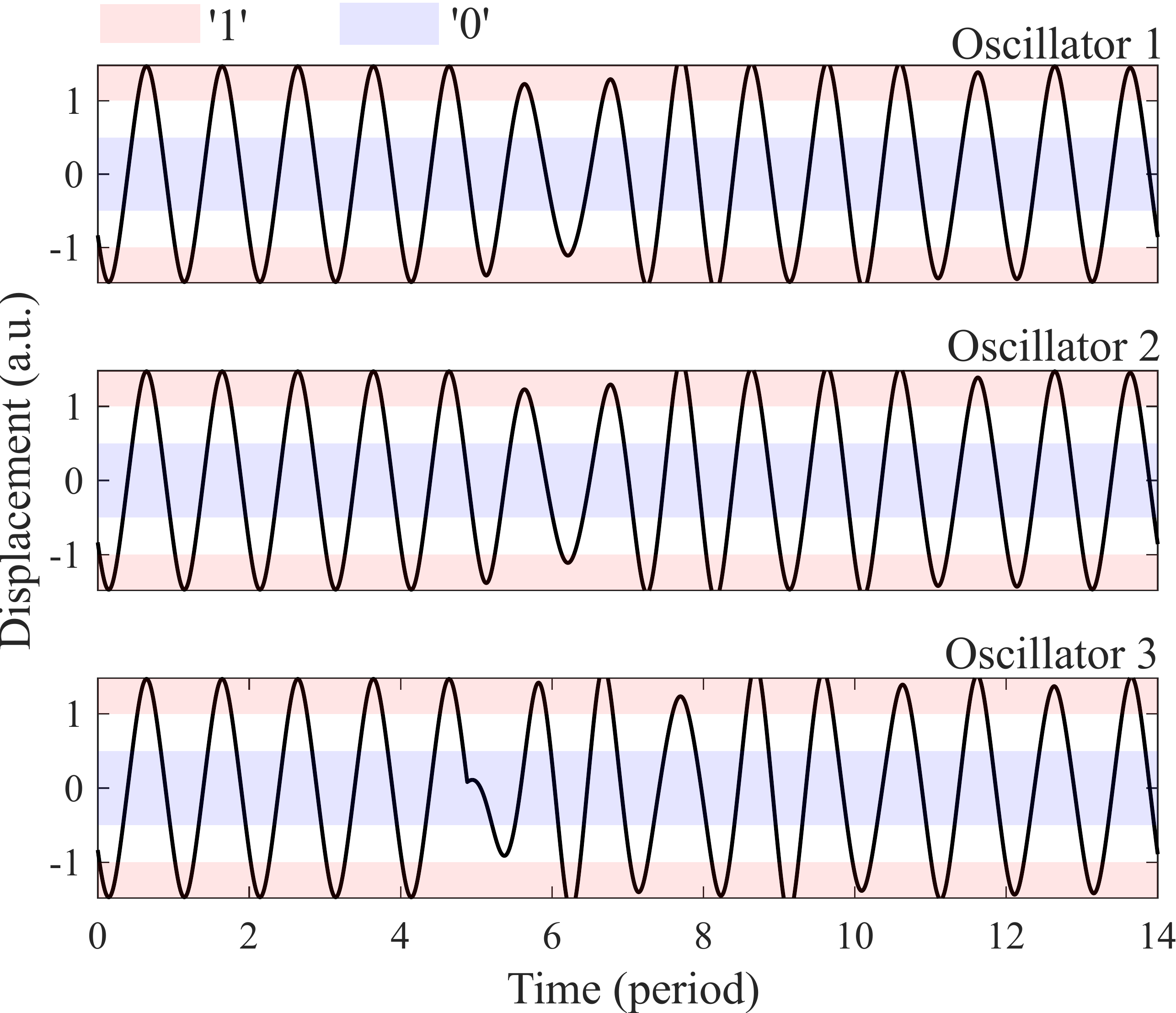}
\end{center}
\caption{Time dynamics of error correction device. The coupled oscillators are initialised in their `1' states and after 4 periods of oscillation, an impulse is applied to the third oscillator. The third oscillator temporarily transitions into its `0' state, but quickly equilibrates back.  The parameters of the oscillator are given by: $m = 10^{-12}$ kg, $\Gamma = 10^{5}$ s$^{-1}$, $\omega_0 = 10^{6}$ s$^{-1}$ , $\alpha = 2 \times 10^{22}$ m$^{-2}.s^{-2}$, $\omega = 1.152 \times 10^{6}$ s$^{-1}$, $F=1.23 \times 10^{-6}$ N, $\Delta p = 6 \times 10^{-12}$ kg.m.s$^{-1}$, $\beta = 2 \times 10^{11}$ s$^{-2}$.
}
\label{Supp Figure 1}
\end{figure}

Choosing system parameters within the error correction regime, the time dynamics of the system exposed to a random impulse is illustrated in Fig.~\ref{Supp Figure 1}b). Each oscillator is initially prepared in the `1' state then oscillator 3 is subjected to an impulse $\delta p$. We see that the impulse causes the amplitude of oscillator 3 to temporarily reach the `0' state, however, it does not latch to this state. After an equilibration period, all oscillators settle back into the `1' state. This represents a successful error correction event.

\section{Simulation of error correction map}
FIG. 2 c) was produced using the following two algorithms:
First, we determined the boundary between the three main regions (`1' bias, `0' bias and retention). To achieve this, the following process was performed:

\begin{enumerate}
\item For a given set of parameters ($\omega_0$, $\alpha$, $\Gamma$ and $\beta$), choose a particular $\omega$ and $F$. 
\item Initialise the oscillators in the `0' states (low displacement) or `1' states (high displacement).
\item Let the oscillators come to equilibrium and record their equilibrium displacement. 
\item If the oscillators equilibrate to a steady state that matches the initial conditions (i.e. '111' or '000'), then the system is in the retention region. Otherwise, if the oscillators equilibrate to their '1' (`0') state regardless of their initial condition, the system is in the `1' (`0') bias region. 
\item Repeat steps 1 - 4 for many different choices of $\omega$ and $F$, to produce a base phase map.
\end{enumerate}

The above steps allowed us to divide the phase space into three regions; `0' bias, `1' bias and retention. The second algorithm was then used to locate the error correction region:

\begin{enumerate}
\item For a given set of parameters ($\omega_0$, $\alpha$, $\Gamma$ and $\beta$), choose a particular $\omega$ and $F$. 
\item Initialise and equilibrate all the oscillators in the `0' states (low displacement) or in the `1' states (high displacement).
\item At a chosen time, introduce an instantaneous kick by changing the momentum of one of the oscillators. The new momentum, $p_1$, is given by $p_0 + \Delta p$, where $p_0$ is the original momentum and $\Delta p$ is the momentum change from the introduced kick. 
\item Let the oscillators come to equilibrium and observe whether the final states of the oscillators match the initial states.
\item Repeat steps 2 - 4 for many different phases of oscillation. If the system corrects errors at all tested error times, designate the set of $\omega$ and $F$ as error correcting.
\item Repeat steps 1 - 5 for many different choices of $\omega$ and $F$, to produce an error correction parameter space map.
\end{enumerate}

This produced an error correction map, showing the region where the system can correct for individual single event upsets (SEUs) introduced at any time. Note that within a cycle of oscillation, the momentum of an oscillator varies. As a result, an oscillator is expected to behave differently when a kick of fixed size $(\Delta p)$ is introduced at different phases in the oscillation. An error correction system must be able to correct errors introduced at any phase. Therefore, we have designed the algorithm to repeat the error correction tests by introducing kicks at different times, corresponding to different phases. Here, we tested 15 linearly spaced phases in our algorithm, with error correction only designated as achieved if the error was corrected at all phases. 

Combining the results from the above two algorithms, the phase map of three coupled Duffing oscillators was determined, as shown in FIG. 2 c).  \\

\section{Simulation (FIG. 4)}

We developed a simulation that is able to repeatedly test whether a system fails to error correct with randomly imposed kicks. Here, we assume that there is an underlying probability of an oscillator experiencing a kick, $P_{kick}$. We have assumed a kick probability of $0.03$ per time interval. Note that given these values, the chance of two or more kicks occurring simultaneously on a single oscillator is negligible. As a result, we have chosen to implement a maximum of one kick per oscillator area in our simulation.\\

The algorithm used to test the performance of a given system (FIG. 4) is shown below:

\begin{enumerate}
\item Choose the parameters of the single and coupled oscillators, and the total number of tests. The coupled oscillators should be in the error correction regime. For fair comparison, the single oscillator should be in its retention region.
\item  Initialise and equilibrate both systems in their `0’ states, record their equilibrium amplitudes. Similarly, initialise both systems in their `1’ states and record their equilibrium amplitudes.
\item Use random number generators to decide the number of oscillators experiencing kicks and impose them onto the system at a randomly generated time. For a single oscillator, one random number generator is used, whereas three random number generators are used for three coupled oscillators. For each generated number, $r \in [0,1]$, a kick is imposed whenever $r < P_{kick}$.
\item Let the system come to equilibrium and record the equilibrium amplitudes. Compare the final and the initial equilibrium amplitudes to determine whether the system has failed. Note that it is considered a failure if either the `0' or `1' states have been altered by the occurrence of kicks.
\item Repeat steps 3 - 4 for the total number of tests. Record the total number of failures. \\
\end{enumerate}

\section{Theoretical distribution of system failure (FIG. 4)}

Theoretically, we can predict the cumulative distribution function (CDF) of system failure for both systems. At a given time window, the CDF is given by the converse of the system succeeding up until that time window. This can be written as:
\begin{align}\label{CDF}
CDF(\tilde{t}) &= 1-(1-P_{fail})^{\tilde{t}}
\end{align}where $P_{fail}$ is the probability of failure for a given system in a given time window. The subscripts $s$ and $c$ will be used to denote single and coupled systems, respectively. Now, we will focus on determining $P_{fail}$ for the two systems of interest.\\

Within our simulation, we assume that all oscillators are of the same area. In addition, we assume that the oscillator areas are sufficiently small such that the probability of an oscillator experiencing multiple kicks is negligible. That is, in a given period of time (time window), each oscillator must experience one of two outcomes; either kicked or not kicked. If we take the probability of a kick per oscillator area per time window as $P_{kick}$, then the probability of failure for a single oscillator in any given time window is given by:
\begin{align} \label{P fail single}
P_{fail_s} = P_{kick}
\end{align}
This is because we set the kick amplitude such that all kick events result in a single oscillator failure.\\

In comparison, we expect three coupled oscillators to function as a three way majority voting system when operating in the error correcting regime. As a result, we anticipate this error correction system to fail only when multiple oscillators experience kicks during the same time window. Therefore, to calculate the probability of multiple oscillators experiencing kicks within the same time window, we can apply the Binomial theorem:
\begin{align} \label{Binomial}
P_{binomial} (j)&= {3\choose j} (1-P_{kick})^{ 3-j} P_{kick}^{j} \end{align} 
where $j$ is the number of oscillators experiencing kicks. Note that $j$ is an integer in the range from 0 to 3. Here, the Binomial theorem is used to account for the different combinations of oscillator outcomes (kicked or not kicked). Using equation \eqref{Binomial}, the predicted probability of failure for three coupled oscillators is given by:\begin{align} \label{P_fail_couple}
P_{fail_c}&= \sum_{j=2}^{3}{3\choose j} (1-P_{kick})^{ 3-j} (P_{kick})^{j} \nonumber
\\
&= 3  (1 - P_{kick}) (P_{kick})^2   + (P_{kick})^3 
\end{align}which is the sum of the probability of the three coupled oscillators experiencing two or three kicks within the same time window. \\

Using equations \eqref{P fail single} and \eqref{P_fail_couple}, we may calculate the values of $P_{fail}$ for a specific $P_{kick}$. For instance, when $P_{kick} = 0.03$, we calculated:\begin{align*}
P_{fail_s} & = 0.03\\
P_{fail_c} & = 0.0026
\end{align*}
Finally, by substituting these numbers into equation \eqref{CDF}, we obtained the theoretical CDFs for both single and coupled oscillators. \\

Similarly, the theoretical prediction of probability of failure per event, $\tilde{P}_{fail}$, as shown in FIG. 4b) can be obtained using equations \eqref{P fail single} and \eqref{P_fail_couple}. Note that the probability of failure presented in FIG. 4b) is normalised to a per event basis. This was achieved by dividing equations \eqref{P fail single} and \eqref{P_fail_couple} by the probability of an event occurring in each scenario:\begin{align*}
\tilde{P}_{fail_s} &= \frac{P_{kick} }{ P_{kick} }\\
& = 1\\
\tilde{P}_{fail_c} &= \frac{3  (1 - P_{kick}) (P_{kick})^2   + (P_{kick})^3 }{ 3 (1 - P_{kick})^2 ( P_{kick}) + 3  (1 - P_{kick}) (P_{kick})^2   + (P_{kick})^3 }
\end{align*}

When using the above equations, we predict a single oscillator to always fail whenever an event occurs, and the coupled systems to only fail whenever two or more kicks are imposed. \\

\section{Difference between theoretical and numerical results (FIG. 4)}

From FIG. 4, it can be observed that the theoretical predictions of the system failure rate are not consistent with the simulated data. Therefore, our previous prediction of $P_{fail}$ must not account for all kicks that the systems can correct. To rectify this, we assume that the single system is capable of correcting $x_s$ proportion of single kicks. Similarly, we assume that the coupled system not only corrects for all single kicks, but also corrects $x_c$ proportion of double kicks. With these assumptions, equations \eqref{P fail single} and \eqref{P_fail_couple} become:
\begin{align} 
P_{fail_s} &= (1 - x_s) P_{kick}\label{corrected models 1}\\
P_{fail_c} &= (1 - x_c) \cdot  (1 - P_{kick}) P_{kick}^2   + P_{kick}^3 \label{corrected models 2}
\end{align}

Using the above expressions, we have performed a non-linear fit to find the values of $x_s$ and $x_c$, that give the best agreement with the simulated data. The fitting was done using the $lsqnonlin$ MATLAB function. This function performs a non-linear least squares fit of the simulation data over a range of kick probabilities to equations \eqref{corrected models 1} - \eqref{corrected models 2}. The results of the non-linear fit (presented in FIG. 4)) are given by:
 \begin{align*}
 x_s &= (8.8 \pm 0.4)\%\\
 x_c &=( 64.0 \pm 0.5)\%
 \end{align*}

This suggests that the single system can correct single kicks approximately $9 \%$ of the time, and the coupled system can correct approximately $64 \%$ of double kicks. With these modified assumptions, the theoretical predictions are observed to agree with the simulated data, as shown in FIG. 4.


\bibliographystyle{apsrev4-2}

\begin{thebibliography}{30}%
\makeatletter
\providecommand \@ifxundefined [1]{%
 \@ifx{#1\undefined}
}%
\providecommand \@ifnum [1]{%
 \ifnum #1\expandafter \@firstoftwo
 \else \expandafter \@secondoftwo
 \fi
}%
\providecommand \@ifx [1]{%
 \ifx #1\expandafter \@firstoftwo
 \else \expandafter \@secondoftwo
 \fi
}%
\providecommand \natexlab [1]{#1}%
\providecommand \enquote  [1]{``#1''}%
\providecommand \bibnamefont  [1]{#1}%
\providecommand \bibfnamefont [1]{#1}%
\providecommand \citenamefont [1]{#1}%
\providecommand \href@noop [0]{\@secondoftwo}%
\providecommand \href [0]{\begingroup \@sanitize@url \@href}%
\providecommand \@href[1]{\@@startlink{#1}\@@href}%
\providecommand \@@href[1]{\endgroup#1\@@endlink}%
\providecommand \@sanitize@url [0]{\catcode `\\12\catcode `\$12\catcode `\&12\catcode `\#12\catcode `\^12\catcode `\_12\catcode `\%12\relax}%
\providecommand \@@startlink[1]{}%
\providecommand \@@endlink[0]{}%
\providecommand \url  [0]{\begingroup\@sanitize@url \@url }%
\providecommand \@url [1]{\endgroup\@href {#1}{\urlprefix }}%
\providecommand \urlprefix  [0]{URL }%
\providecommand \Eprint [0]{\href }%
\providecommand \doibase [0]{https://doi.org/}%
\providecommand \selectlanguage [0]{\@gobble}%
\providecommand \bibinfo  [0]{\@secondoftwo}%
\providecommand \bibfield  [0]{\@secondoftwo}%
\providecommand \translation [1]{[#1]}%
\providecommand \BibitemOpen [0]{}%
\providecommand \bibitemStop [0]{}%
\providecommand \bibitemNoStop [0]{.\EOS\space}%
\providecommand \EOS [0]{\spacefactor3000\relax}%
\providecommand \BibitemShut  [1]{\csname bibitem#1\endcsname}%
\let\auto@bib@innerbib\@empty
\bibitem [{\citenamefont {Restrepo}\ \emph {et~al.}(2005)\citenamefont {Restrepo}, \citenamefont {Ott},\ and\ \citenamefont {Hunt}}]{Restrepo_PRE_05}%
  \BibitemOpen
  \bibfield  {author} {\bibinfo {author} {\bibfnamefont {J.~G.}\ \bibnamefont {Restrepo}}, \bibinfo {author} {\bibfnamefont {E.}~\bibnamefont {Ott}},\ and\ \bibinfo {author} {\bibfnamefont {B.~R.}\ \bibnamefont {Hunt}},\ }\href {https://doi.org/10.1103/PhysRevE.71.036151} {\bibfield  {journal} {\bibinfo  {journal} {Phys. Rev. E}\ }\textbf {\bibinfo {volume} {71}},\ \bibinfo {pages} {036151} (\bibinfo {year} {2005})}\BibitemShut {NoStop}%
\bibitem [{\citenamefont {Abrams}\ and\ \citenamefont {Strogatz}(2004)}]{Abrams_PRL_04}%
  \BibitemOpen
  \bibfield  {author} {\bibinfo {author} {\bibfnamefont {D.~M.}\ \bibnamefont {Abrams}}\ and\ \bibinfo {author} {\bibfnamefont {S.~H.}\ \bibnamefont {Strogatz}},\ }\href {https://doi.org/10.1103/PhysRevLett.93.174102} {\bibfield  {journal} {\bibinfo  {journal} {Phys. Rev. Lett.}\ }\textbf {\bibinfo {volume} {93}},\ \bibinfo {pages} {174102} (\bibinfo {year} {2004})}\BibitemShut {NoStop}%
\bibitem [{\citenamefont {Sarfati}\ and\ \citenamefont {Peleg}(2022)}]{Sarfati_SciAdv_22}%
  \BibitemOpen
  \bibfield  {author} {\bibinfo {author} {\bibfnamefont {R.}~\bibnamefont {Sarfati}}\ and\ \bibinfo {author} {\bibfnamefont {O.}~\bibnamefont {Peleg}},\ }\href {https://doi.org/10.1126/sciadv.add6690} {\bibfield  {journal} {\bibinfo  {journal} {Science Advances}\ }\textbf {\bibinfo {volume} {8}},\ \bibinfo {pages} {eadd6690} (\bibinfo {year} {2022})}\BibitemShut {NoStop}%
\bibitem [{\citenamefont {Uhlhaas}\ and\ \citenamefont {Singer}(2006)}]{Uhlhass_Neuron_06}%
  \BibitemOpen
  \bibfield  {author} {\bibinfo {author} {\bibfnamefont {P.~J.}\ \bibnamefont {Uhlhaas}}\ and\ \bibinfo {author} {\bibfnamefont {W.}~\bibnamefont {Singer}},\ }\href {https://doi.org/10.1016/j.neuron.2006.09.020} {\bibfield  {journal} {\bibinfo  {journal} {Neuron}\ }\textbf {\bibinfo {volume} {52}},\ \bibinfo {pages} {155} (\bibinfo {year} {2006})}\BibitemShut {NoStop}%
\bibitem [{\citenamefont {Sakurai}\ \emph {et~al.}(2021)\citenamefont {Sakurai}, \citenamefont {Bastidas}, \citenamefont {Munro},\ and\ \citenamefont {Nemoto}}]{Sakurai_PRL_21}%
  \BibitemOpen
  \bibfield  {author} {\bibinfo {author} {\bibfnamefont {A.}~\bibnamefont {Sakurai}}, \bibinfo {author} {\bibfnamefont {V.~M.}\ \bibnamefont {Bastidas}}, \bibinfo {author} {\bibfnamefont {W.~J.}\ \bibnamefont {Munro}},\ and\ \bibinfo {author} {\bibfnamefont {K.}~\bibnamefont {Nemoto}},\ }\href {https://doi.org/10.1103/PhysRevLett.126.120606} {\bibfield  {journal} {\bibinfo  {journal} {Phys. Rev. Lett.}\ }\textbf {\bibinfo {volume} {126}},\ \bibinfo {pages} {120606} (\bibinfo {year} {2021})}\BibitemShut {NoStop}%
\bibitem [{\citenamefont {K{\'o}m{\'a}r}\ \emph {et~al.}(2014)\citenamefont {K{\'o}m{\'a}r}, \citenamefont {Kessler}, \citenamefont {Bishof}, \citenamefont {Jiang}, \citenamefont {S{\o}rensen}, \citenamefont {Ye},\ and\ \citenamefont {Lukin}}]{Komar_NatPhys_2014}%
  \BibitemOpen
  \bibfield  {author} {\bibinfo {author} {\bibfnamefont {P.}~\bibnamefont {K{\'o}m{\'a}r}}, \bibinfo {author} {\bibfnamefont {E.~M.}\ \bibnamefont {Kessler}}, \bibinfo {author} {\bibfnamefont {M.}~\bibnamefont {Bishof}}, \bibinfo {author} {\bibfnamefont {L.}~\bibnamefont {Jiang}}, \bibinfo {author} {\bibfnamefont {A.~S.}\ \bibnamefont {S{\o}rensen}}, \bibinfo {author} {\bibfnamefont {J.}~\bibnamefont {Ye}},\ and\ \bibinfo {author} {\bibfnamefont {M.~D.}\ \bibnamefont {Lukin}},\ }\href@noop {} {\bibfield  {journal} {\bibinfo  {journal} {Nature Physics}\ }\textbf {\bibinfo {volume} {10}},\ \bibinfo {pages} {582} (\bibinfo {year} {2014})}\BibitemShut {NoStop}%
\bibitem [{\citenamefont {Torrejon}\ \emph {et~al.}(2017)\citenamefont {Torrejon}, \citenamefont {Riou}, \citenamefont {Araujo}, \citenamefont {Tsunegi}, \citenamefont {Khalsa}, \citenamefont {Querlioz}, \citenamefont {Bortolotti}, \citenamefont {Cros}, \citenamefont {Yakushiji}, \citenamefont {Fukushima}, \citenamefont {Kubota}, \citenamefont {Yuasa}, \citenamefont {Stiles},\ and\ \citenamefont {Grollier}}]{Torrejon_Nat_2017}%
  \BibitemOpen
  \bibfield  {author} {\bibinfo {author} {\bibfnamefont {J.}~\bibnamefont {Torrejon}}, \bibinfo {author} {\bibfnamefont {M.}~\bibnamefont {Riou}}, \bibinfo {author} {\bibfnamefont {F.~A.}\ \bibnamefont {Araujo}}, \bibinfo {author} {\bibfnamefont {S.}~\bibnamefont {Tsunegi}}, \bibinfo {author} {\bibfnamefont {G.}~\bibnamefont {Khalsa}}, \bibinfo {author} {\bibfnamefont {D.}~\bibnamefont {Querlioz}}, \bibinfo {author} {\bibfnamefont {P.}~\bibnamefont {Bortolotti}}, \bibinfo {author} {\bibfnamefont {V.}~\bibnamefont {Cros}}, \bibinfo {author} {\bibfnamefont {K.}~\bibnamefont {Yakushiji}}, \bibinfo {author} {\bibfnamefont {A.}~\bibnamefont {Fukushima}}, \bibinfo {author} {\bibfnamefont {H.}~\bibnamefont {Kubota}}, \bibinfo {author} {\bibfnamefont {S.}~\bibnamefont {Yuasa}}, \bibinfo {author} {\bibfnamefont {M.~D.}\ \bibnamefont {Stiles}},\ and\ \bibinfo {author} {\bibfnamefont {J.}~\bibnamefont {Grollier}},\ }\href@noop {} {\bibfield  {journal} {\bibinfo  {journal} {Nature}\ }\textbf {\bibinfo {volume} {547}},\
  \bibinfo {pages} {428} (\bibinfo {year} {2017})}\BibitemShut {NoStop}%
\bibitem [{\citenamefont {Xu}\ \emph {et~al.}(2020)\citenamefont {Xu}, \citenamefont {Wang}, \citenamefont {Jiang},\ and\ \citenamefont {Wei}}]{Xu_MicroNano_2020}%
  \BibitemOpen
  \bibfield  {author} {\bibinfo {author} {\bibfnamefont {L.}~\bibnamefont {Xu}}, \bibinfo {author} {\bibfnamefont {S.}~\bibnamefont {Wang}}, \bibinfo {author} {\bibfnamefont {Z.}~\bibnamefont {Jiang}},\ and\ \bibinfo {author} {\bibfnamefont {X.}~\bibnamefont {Wei}},\ }\href@noop {} {\bibfield  {journal} {\bibinfo  {journal} {Microsystems \& Nanoengineering}\ }\textbf {\bibinfo {volume} {6}},\ \bibinfo {pages} {63} (\bibinfo {year} {2020})}\BibitemShut {NoStop}%
\bibitem [{\citenamefont {Boaknin}\ \emph {et~al.}(2007)\citenamefont {Boaknin}, \citenamefont {Manucharyan}, \citenamefont {Fissette}, \citenamefont {Metcalfe}, \citenamefont {Frunzio}, \citenamefont {Vijay}, \citenamefont {Siddiqi}, \citenamefont {Wallraff}, \citenamefont {Schoelkopf},\ and\ \citenamefont {Devoret}}]{Boaknin}%
  \BibitemOpen
  \bibfield  {author} {\bibinfo {author} {\bibfnamefont {E.}~\bibnamefont {Boaknin}}, \bibinfo {author} {\bibfnamefont {V.~E.}\ \bibnamefont {Manucharyan}}, \bibinfo {author} {\bibfnamefont {S.}~\bibnamefont {Fissette}}, \bibinfo {author} {\bibfnamefont {M.}~\bibnamefont {Metcalfe}}, \bibinfo {author} {\bibfnamefont {L.}~\bibnamefont {Frunzio}}, \bibinfo {author} {\bibfnamefont {R.}~\bibnamefont {Vijay}}, \bibinfo {author} {\bibfnamefont {I.}~\bibnamefont {Siddiqi}}, \bibinfo {author} {\bibfnamefont {A.}~\bibnamefont {Wallraff}}, \bibinfo {author} {\bibfnamefont {R.~J.}\ \bibnamefont {Schoelkopf}},\ and\ \bibinfo {author} {\bibfnamefont {M.}~\bibnamefont {Devoret}},\ }\href@noop {} {\bibinfo {title} {Dispersive microwave bifurcation of a superconducting resonator cavity incorporating a josephson junction}} (\bibinfo {year} {2007}),\ \Eprint {https://arxiv.org/abs/cond-mat/0702445} {arXiv:cond-mat/0702445 [cond-mat.supr-con]} \BibitemShut {NoStop}%
\bibitem [{\citenamefont {MOSS}\ and\ \citenamefont {EGGLETON}(2008)}]{Moss}%
  \BibitemOpen
  \bibfield  {author} {\bibinfo {author} {\bibfnamefont {D.~J.}\ \bibnamefont {MOSS}}\ and\ \bibinfo {author} {\bibfnamefont {B.~J.}\ \bibnamefont {EGGLETON}},\ }\href@noop {} {\emph {\bibinfo {title} {Advances in Information Optics and Photonics}}},\ Vol.~\bibinfo {volume} {6}\ (\bibinfo  {publisher} {SPIE},\ \bibinfo {year} {2008})\ pp.\ \bibinfo {pages} {657--686}\BibitemShut {NoStop}%
\bibitem [{\citenamefont {Dreyer}\ \emph {et~al.}(2022)\citenamefont {Dreyer}, \citenamefont {Schäffer}, \citenamefont {Bauer}, \citenamefont {Liebing}, \citenamefont {Berakdar},\ and\ \citenamefont {Woltersdorf}}]{Woltersdorf_NatCom_2022}%
  \BibitemOpen
  \bibfield  {author} {\bibinfo {author} {\bibfnamefont {R.}~\bibnamefont {Dreyer}}, \bibinfo {author} {\bibfnamefont {A.~F.}\ \bibnamefont {Schäffer}}, \bibinfo {author} {\bibfnamefont {H.~G.}\ \bibnamefont {Bauer}}, \bibinfo {author} {\bibfnamefont {N.}~\bibnamefont {Liebing}}, \bibinfo {author} {\bibfnamefont {J.}~\bibnamefont {Berakdar}},\ and\ \bibinfo {author} {\bibfnamefont {G.}~\bibnamefont {Woltersdorf}},\ }\href {https://doi.org/10.1038/s41467-022-32224-0} {\bibfield  {journal} {\bibinfo  {journal} {Nature Communications}\ }\textbf {\bibinfo {volume} {13}},\ \bibinfo {pages} {4939} (\bibinfo {year} {2022})}\BibitemShut {NoStop}%
\bibitem [{\citenamefont {Gavartin}\ \emph {et~al.}(2013)\citenamefont {Gavartin}, \citenamefont {Verlot},\ and\ \citenamefont {Kippenberg}}]{gavartin2013stabilization}%
  \BibitemOpen
  \bibfield  {author} {\bibinfo {author} {\bibfnamefont {E.}~\bibnamefont {Gavartin}}, \bibinfo {author} {\bibfnamefont {P.}~\bibnamefont {Verlot}},\ and\ \bibinfo {author} {\bibfnamefont {T.~J.}\ \bibnamefont {Kippenberg}},\ }\href@noop {} {\bibfield  {journal} {\bibinfo  {journal} {Nature communications}\ }\textbf {\bibinfo {volume} {4}},\ \bibinfo {pages} {1} (\bibinfo {year} {2013})}\BibitemShut {NoStop}%
\bibitem [{\citenamefont {Zhang}\ and\ \citenamefont {Turner}(2005)}]{zhang2005application}%
  \BibitemOpen
  \bibfield  {author} {\bibinfo {author} {\bibfnamefont {W.}~\bibnamefont {Zhang}}\ and\ \bibinfo {author} {\bibfnamefont {K.~L.}\ \bibnamefont {Turner}},\ }\href@noop {} {\bibfield  {journal} {\bibinfo  {journal} {Sensors and Actuators A: Physical}\ }\textbf {\bibinfo {volume} {122}},\ \bibinfo {pages} {23} (\bibinfo {year} {2005})}\BibitemShut {NoStop}%
\bibitem [{\citenamefont {Delsing}\ \emph {et~al.}(2019)\citenamefont {Delsing}, \citenamefont {Cleland}, \citenamefont {Schuetz}, \citenamefont {Kn{\"o}rzer}, \citenamefont {Giedke}, \citenamefont {Cirac}, \citenamefont {Srinivasan}, \citenamefont {Wu}, \citenamefont {Balram}, \citenamefont {B{\"a}uerle} \emph {et~al.}}]{delsing_JPhysD_2019}%
  \BibitemOpen
  \bibfield  {author} {\bibinfo {author} {\bibfnamefont {P.}~\bibnamefont {Delsing}}, \bibinfo {author} {\bibfnamefont {A.~N.}\ \bibnamefont {Cleland}}, \bibinfo {author} {\bibfnamefont {M.~J.}\ \bibnamefont {Schuetz}}, \bibinfo {author} {\bibfnamefont {J.}~\bibnamefont {Kn{\"o}rzer}}, \bibinfo {author} {\bibfnamefont {G.}~\bibnamefont {Giedke}}, \bibinfo {author} {\bibfnamefont {J.~I.}\ \bibnamefont {Cirac}}, \bibinfo {author} {\bibfnamefont {K.}~\bibnamefont {Srinivasan}}, \bibinfo {author} {\bibfnamefont {M.}~\bibnamefont {Wu}}, \bibinfo {author} {\bibfnamefont {K.~C.}\ \bibnamefont {Balram}}, \bibinfo {author} {\bibfnamefont {C.}~\bibnamefont {B{\"a}uerle}}, \emph {et~al.},\ }\href@noop {} {\bibfield  {journal} {\bibinfo  {journal} {Journal of Physics D: Applied Physics}\ }\textbf {\bibinfo {volume} {52}},\ \bibinfo {pages} {353001} (\bibinfo {year} {2019})}\BibitemShut {NoStop}%
\bibitem [{\citenamefont {R{\"u}tzel}\ \emph {et~al.}(2003)\citenamefont {R{\"u}tzel}, \citenamefont {Lee},\ and\ \citenamefont {Raman}}]{rutzel_RoyalSoc_2003}%
  \BibitemOpen
  \bibfield  {author} {\bibinfo {author} {\bibfnamefont {S.}~\bibnamefont {R{\"u}tzel}}, \bibinfo {author} {\bibfnamefont {S.~I.}\ \bibnamefont {Lee}},\ and\ \bibinfo {author} {\bibfnamefont {A.}~\bibnamefont {Raman}},\ }\href@noop {} {\bibfield  {journal} {\bibinfo  {journal} {Proceedings of the Royal Society of London. Series A: Mathematical, Physical and Engineering Sciences}\ }\textbf {\bibinfo {volume} {459}},\ \bibinfo {pages} {1925} (\bibinfo {year} {2003})}\BibitemShut {NoStop}%
\bibitem [{\citenamefont {Coulombe}\ \emph {et~al.}(2017)\citenamefont {Coulombe}, \citenamefont {York},\ and\ \citenamefont {Sylvestre}}]{Coulombe_PLOSONE_2017}%
  \BibitemOpen
  \bibfield  {author} {\bibinfo {author} {\bibfnamefont {J.~C.}\ \bibnamefont {Coulombe}}, \bibinfo {author} {\bibfnamefont {M.~C.~A.}\ \bibnamefont {York}},\ and\ \bibinfo {author} {\bibfnamefont {J.}~\bibnamefont {Sylvestre}},\ }\href {https://doi.org/10.1371/journal.pone.0178663} {\bibfield  {journal} {\bibinfo  {journal} {PLOS ONE}\ }\textbf {\bibinfo {volume} {12}},\ \bibinfo {pages} {1} (\bibinfo {year} {2017})}\BibitemShut {NoStop}%
\bibitem [{\citenamefont {Song}\ \emph {et~al.}(2019)\citenamefont {Song}, \citenamefont {Panas}, \citenamefont {Chizari}, \citenamefont {Shaw}, \citenamefont {Jackson}, \citenamefont {Hopkins},\ and\ \citenamefont {Pascall}}]{Song_19}%
  \BibitemOpen
  \bibfield  {author} {\bibinfo {author} {\bibfnamefont {Y.}~\bibnamefont {Song}}, \bibinfo {author} {\bibfnamefont {R.~M.}\ \bibnamefont {Panas}}, \bibinfo {author} {\bibfnamefont {S.}~\bibnamefont {Chizari}}, \bibinfo {author} {\bibfnamefont {L.~A.}\ \bibnamefont {Shaw}}, \bibinfo {author} {\bibfnamefont {J.~A.}\ \bibnamefont {Jackson}}, \bibinfo {author} {\bibfnamefont {J.~B.}\ \bibnamefont {Hopkins}},\ and\ \bibinfo {author} {\bibfnamefont {A.~J.}\ \bibnamefont {Pascall}},\ }\href {https://doi.org/10.1038/s41467-019-08678-0} {\bibfield  {journal} {\bibinfo  {journal} {Nature communications}\ }\textbf {\bibinfo {volume} {10}},\ \bibinfo {pages} {882} (\bibinfo {year} {2019})}\BibitemShut {NoStop}%
\bibitem [{\citenamefont {Wenzler}\ \emph {et~al.}(2014)\citenamefont {Wenzler}, \citenamefont {Dunn}, \citenamefont {Toffoli},\ and\ \citenamefont {Mohanty}}]{Wenzler}%
  \BibitemOpen
  \bibfield  {author} {\bibinfo {author} {\bibfnamefont {J.-S.}\ \bibnamefont {Wenzler}}, \bibinfo {author} {\bibfnamefont {T.}~\bibnamefont {Dunn}}, \bibinfo {author} {\bibfnamefont {T.}~\bibnamefont {Toffoli}},\ and\ \bibinfo {author} {\bibfnamefont {P.}~\bibnamefont {Mohanty}},\ }\href@noop {} {\bibfield  {journal} {\bibinfo  {journal} {Nano Lett.}\ }\textbf {\bibinfo {volume} {14}},\ \bibinfo {pages} {89} (\bibinfo {year} {2014})}\BibitemShut {NoStop}%
\bibitem [{\citenamefont {Yamaguchi}\ \emph {et~al.}(2011)\citenamefont {Yamaguchi}, \citenamefont {Nishiguchi}, \citenamefont {Flurin}, \citenamefont {Fujiwara},\ and\ \citenamefont {Mahboob}}]{Yamaguchi_11}%
  \BibitemOpen
  \bibfield  {author} {\bibinfo {author} {\bibfnamefont {H.}~\bibnamefont {Yamaguchi}}, \bibinfo {author} {\bibfnamefont {K.}~\bibnamefont {Nishiguchi}}, \bibinfo {author} {\bibfnamefont {E.}~\bibnamefont {Flurin}}, \bibinfo {author} {\bibfnamefont {A.}~\bibnamefont {Fujiwara}},\ and\ \bibinfo {author} {\bibfnamefont {I.}~\bibnamefont {Mahboob}},\ }\href {https://doi.org/10.1038/ncomms1201} {\bibfield  {journal} {\bibinfo  {journal} {Nature communications}\ }\textbf {\bibinfo {volume} {2}},\ \bibinfo {pages} {198} (\bibinfo {year} {2011})}\BibitemShut {NoStop}%
\bibitem [{\citenamefont {Yao}\ and\ \citenamefont {Hikihara}(2014{\natexlab{a}})}]{Yao_14}%
  \BibitemOpen
  \bibfield  {author} {\bibinfo {author} {\bibfnamefont {A.}~\bibnamefont {Yao}}\ and\ \bibinfo {author} {\bibfnamefont {T.}~\bibnamefont {Hikihara}},\ }\href {https://doi.org/10.1063/1.4896272} {\bibfield  {journal} {\bibinfo  {journal} {Applied physics letters}\ }\textbf {\bibinfo {volume} {105}},\ \bibinfo {pages} {123104} (\bibinfo {year} {2014}{\natexlab{a}})}\BibitemShut {NoStop}%
\bibitem [{\citenamefont {Roukes}(2004)}]{Roukes}%
  \BibitemOpen
  \bibfield  {author} {\bibinfo {author} {\bibfnamefont {M.}~\bibnamefont {Roukes}},\ }in\ \href@noop {} {\emph {\bibinfo {booktitle} {IEDM Technical Digest. IEEE International Electron Devices Meeting, 2004}}}\ (\bibinfo  {publisher} {IEEE},\ \bibinfo {year} {2004})\ pp.\ \bibinfo {pages} {539--542}\BibitemShut {NoStop}%
\bibitem [{\citenamefont {Romero}\ \emph {et~al.}(2022)\citenamefont {Romero}, \citenamefont {Mauranyapin}, \citenamefont {Hirsch}, \citenamefont {Kalra}, \citenamefont {Baker}, \citenamefont {Harris},\ and\ \citenamefont {Bowen}}]{Romero_arxiv22}%
  \BibitemOpen
  \bibfield  {author} {\bibinfo {author} {\bibfnamefont {E.}~\bibnamefont {Romero}}, \bibinfo {author} {\bibfnamefont {N.~P.}\ \bibnamefont {Mauranyapin}}, \bibinfo {author} {\bibfnamefont {T.~M.~F.}\ \bibnamefont {Hirsch}}, \bibinfo {author} {\bibfnamefont {R.}~\bibnamefont {Kalra}}, \bibinfo {author} {\bibfnamefont {C.~G.}\ \bibnamefont {Baker}}, \bibinfo {author} {\bibfnamefont {G.~I.}\ \bibnamefont {Harris}},\ and\ \bibinfo {author} {\bibfnamefont {W.~P.}\ \bibnamefont {Bowen}},\ }\href@noop {} {\bibinfo {title} {Scalable nanomechanical logic gate}} (\bibinfo {year} {2022}),\ \Eprint {https://arxiv.org/abs/cond-mat/2206.11661v2} {arXiv:cond-mat/2206.11661v2 [cond-mat.mes-hall]} \BibitemShut {NoStop}%
\bibitem [{\citenamefont {Lee}\ \emph {et~al.}(2023)\citenamefont {Lee}, \citenamefont {Kang}, \citenamefont {Choi}, \citenamefont {Kim}, \citenamefont {Kim}, \citenamefont {Lee},\ and\ \citenamefont {Yoon}}]{Lee_NatComm_23}%
  \BibitemOpen
  \bibfield  {author} {\bibinfo {author} {\bibfnamefont {Y.-B.}\ \bibnamefont {Lee}}, \bibinfo {author} {\bibfnamefont {M.-H.}\ \bibnamefont {Kang}}, \bibinfo {author} {\bibfnamefont {P.-K.}\ \bibnamefont {Choi}}, \bibinfo {author} {\bibfnamefont {S.-H.}\ \bibnamefont {Kim}}, \bibinfo {author} {\bibfnamefont {T.-S.}\ \bibnamefont {Kim}}, \bibinfo {author} {\bibfnamefont {S.-Y.}\ \bibnamefont {Lee}},\ and\ \bibinfo {author} {\bibfnamefont {J.-B.}\ \bibnamefont {Yoon}},\ }\href@noop {} {\bibfield  {journal} {\bibinfo  {journal} {Nature Communications}\ }\textbf {\bibinfo {volume} {14}},\ \bibinfo {pages} {460} (\bibinfo {year} {2023})}\BibitemShut {NoStop}%
\bibitem [{\citenamefont {Schroeder}\ \emph {et~al.}(2009)\citenamefont {Schroeder}, \citenamefont {Pinheiro},\ and\ \citenamefont {Weber}}]{schroeder_Google_2009}%
  \BibitemOpen
  \bibfield  {author} {\bibinfo {author} {\bibfnamefont {B.}~\bibnamefont {Schroeder}}, \bibinfo {author} {\bibfnamefont {E.}~\bibnamefont {Pinheiro}},\ and\ \bibinfo {author} {\bibfnamefont {W.-D.}\ \bibnamefont {Weber}},\ }\href@noop {} {\bibfield  {journal} {\bibinfo  {journal} {ACM SIGMETRICS Performance Evaluation Review}\ }\textbf {\bibinfo {volume} {37}},\ \bibinfo {pages} {193} (\bibinfo {year} {2009})}\BibitemShut {NoStop}%
\bibitem [{\citenamefont {Karnik}\ and\ \citenamefont {Hazucha}(2004)}]{Karnik_IEEE_04}%
  \BibitemOpen
  \bibfield  {author} {\bibinfo {author} {\bibfnamefont {T.}~\bibnamefont {Karnik}}\ and\ \bibinfo {author} {\bibfnamefont {P.}~\bibnamefont {Hazucha}},\ }\href {https://doi.org/10.1109/TDSC.2004.14} {\bibfield  {journal} {\bibinfo  {journal} {IEEE Transactions on Dependable and Secure Computing}\ }\textbf {\bibinfo {volume} {1}},\ \bibinfo {pages} {128} (\bibinfo {year} {2004})}\BibitemShut {NoStop}%
\bibitem [{\citenamefont {Ziegler}\ and\ \citenamefont {Lanford}(1979)}]{Ziegler_Science_79}%
  \BibitemOpen
  \bibfield  {author} {\bibinfo {author} {\bibfnamefont {J.~F.}\ \bibnamefont {Ziegler}}\ and\ \bibinfo {author} {\bibfnamefont {W.~A.}\ \bibnamefont {Lanford}},\ }\href {https://doi.org/10.1126/science.206.4420.776} {\bibfield  {journal} {\bibinfo  {journal} {Science}\ }\textbf {\bibinfo {volume} {206}},\ \bibinfo {pages} {776} (\bibinfo {year} {1979})}\BibitemShut {NoStop}%
\bibitem [{\citenamefont {Yao}\ and\ \citenamefont {Hikihara}(2014{\natexlab{b}})}]{Yao}%
  \BibitemOpen
  \bibfield  {author} {\bibinfo {author} {\bibfnamefont {A.}~\bibnamefont {Yao}}\ and\ \bibinfo {author} {\bibfnamefont {T.}~\bibnamefont {Hikihara}},\ }\href@noop {} {\bibfield  {journal} {\bibinfo  {journal} {Applied physics letters}\ }\textbf {\bibinfo {volume} {105}},\ \bibinfo {pages} {123104} (\bibinfo {year} {2014}{\natexlab{b}})}\BibitemShut {NoStop}%
\bibitem [{\citenamefont {Tadokoro}\ and\ \citenamefont {Tanaka}(2021)}]{Tadokoro21}%
  \BibitemOpen
  \bibfield  {author} {\bibinfo {author} {\bibfnamefont {Y.}~\bibnamefont {Tadokoro}}\ and\ \bibinfo {author} {\bibfnamefont {H.}~\bibnamefont {Tanaka}},\ }\href {https://doi.org/10.1103/PhysRevApplied.15.024058} {\bibfield  {journal} {\bibinfo  {journal} {Phys. Rev. Appl.}\ }\textbf {\bibinfo {volume} {15}},\ \bibinfo {pages} {024058} (\bibinfo {year} {2021})}\BibitemShut {NoStop}%
\bibitem [{\citenamefont {Schmid}\ \emph {et~al.}(2016)\citenamefont {Schmid}, \citenamefont {Villanueva},\ and\ \citenamefont {Roukes}}]{Schmid}%
  \BibitemOpen
  \bibfield  {author} {\bibinfo {author} {\bibfnamefont {S.}~\bibnamefont {Schmid}}, \bibinfo {author} {\bibfnamefont {L.~G.}\ \bibnamefont {Villanueva}},\ and\ \bibinfo {author} {\bibfnamefont {M.~L.}\ \bibnamefont {Roukes}},\ }\href@noop {} {\emph {\bibinfo {title} {Fundamentals of Nanomechanical Resonators}}}\ (\bibinfo  {publisher} {Springer International Publishing : Imprint: Springer},\ \bibinfo {year} {2016})\BibitemShut {NoStop}%
\bibitem [{\citenamefont {She}\ and\ \citenamefont {McElvain}(2009)}]{TMR}%
  \BibitemOpen
  \bibfield  {author} {\bibinfo {author} {\bibfnamefont {X.}~\bibnamefont {She}}\ and\ \bibinfo {author} {\bibfnamefont {K.}~\bibnamefont {McElvain}},\ }\href@noop {} {\bibfield  {journal} {\bibinfo  {journal} {IEEE transactions on nuclear science}\ }\textbf {\bibinfo {volume} {56}},\ \bibinfo {pages} {2443} (\bibinfo {year} {2009})}\BibitemShut {NoStop}%
\end{thebibliography}

\begin{thebibliography}{2}%
\makeatletter
\providecommand \@ifxundefined [1]{%
 \@ifx{#1\undefined}
}%
\providecommand \@ifnum [1]{%
 \ifnum #1\expandafter \@firstoftwo
 \else \expandafter \@secondoftwo
 \fi
}%
\providecommand \@ifx [1]{%
 \ifx #1\expandafter \@firstoftwo
 \else \expandafter \@secondoftwo
 \fi
}%
\providecommand \natexlab [1]{#1}%
\providecommand \enquote  [1]{``#1''}%
\providecommand \bibnamefont  [1]{#1}%
\providecommand \bibfnamefont [1]{#1}%
\providecommand \citenamefont [1]{#1}%
\providecommand \href@noop [0]{\@secondoftwo}%
\providecommand \href [0]{\begingroup \@sanitize@url \@href}%
\providecommand \@href[1]{\@@startlink{#1}\@@href}%
\providecommand \@@href[1]{\endgroup#1\@@endlink}%
\providecommand \@sanitize@url [0]{\catcode `\\12\catcode `\$12\catcode `\&12\catcode `\#12\catcode `\^12\catcode `\_12\catcode `\%12\relax}%
\providecommand \@@startlink[1]{}%
\providecommand \@@endlink[0]{}%
\providecommand \url  [0]{\begingroup\@sanitize@url \@url }%
\providecommand \@url [1]{\endgroup\@href {#1}{\urlprefix }}%
\providecommand \urlprefix  [0]{URL }%
\providecommand \Eprint [0]{\href }%
\providecommand \doibase [0]{https://doi.org/}%
\providecommand \selectlanguage [0]{\@gobble}%
\providecommand \bibinfo  [0]{\@secondoftwo}%
\providecommand \bibfield  [0]{\@secondoftwo}%
\providecommand \translation [1]{[#1]}%
\providecommand \BibitemOpen [0]{}%
\providecommand \bibitemStop [0]{}%
\providecommand \bibitemNoStop [0]{.\EOS\space}%
\providecommand \EOS [0]{\spacefactor3000\relax}%
\providecommand \BibitemShut  [1]{\csname bibitem#1\endcsname}%
\let\auto@bib@innerbib\@empty
\bibitem [{\citenamefont {Wawrzynski}(2021)}]{Wawrzynski_2021}%
  \BibitemOpen
  \bibfield  {author} {\bibinfo {author} {\bibfnamefont {W.}~\bibnamefont {Wawrzynski}},\ }\href@noop {} {\bibfield  {journal} {\bibinfo  {journal} {Scientific reports}\ }\textbf {\bibinfo {volume} {11}},\ \bibinfo {pages} {2889} (\bibinfo {year} {2021})}\BibitemShut {NoStop}%
\bibitem [{\citenamefont {Brennan}\ \emph {et~al.}(2008)\citenamefont {Brennan}, \citenamefont {Kovacic}, \citenamefont {Carrella},\ and\ \citenamefont {Waters}}]{BrennanM_2018}%
  \BibitemOpen
  \bibfield  {author} {\bibinfo {author} {\bibfnamefont {M.}~\bibnamefont {Brennan}}, \bibinfo {author} {\bibfnamefont {I.}~\bibnamefont {Kovacic}}, \bibinfo {author} {\bibfnamefont {A.}~\bibnamefont {Carrella}},\ and\ \bibinfo {author} {\bibfnamefont {T.}~\bibnamefont {Waters}},\ }\href@noop {} {\bibfield  {journal} {\bibinfo  {journal} {Journal of sound and vibration}\ }\textbf {\bibinfo {volume} {318}},\ \bibinfo {pages} {1250} (\bibinfo {year} {2008})}\BibitemShut {NoStop}%
\end{thebibliography}

\providecommand{\noopsort}[1]{}\providecommand{\singleletter}[1]{#1}%

\end{document}